# Venus as a Nearby Exoplanetary Laboratory

Stephen R. Kane (UC Riverside), Phone: 951-827-6593, Email: skane@ucr.edu

Co-authors: Giada Arney (NASA GSFC), Paul Byrne (NC State University), David Crisp (JPL), Shawn Domagal-Goldman (NASA GSFC), Colin Goldblatt (University of Victoria), David Grinspoon (PSI), James W. Head (Brown University), Adrian Lenardic (Rice University), Victoria Meadows (University of Washington), Cayman Unterborn (Arizona State University), Michael J. Way (NASA GISS)

Co-signers: William Moore (Hampton), Noam Izenberg (APL), Chuanfei Dong (Princeton), Jennifer Whitten (Tulane), Constantine Tsang (SWRI), Martha Gilmore (Wesleyan), Allan Treiman (LPI), Jorn Helbert (IPR), Patrick McGovern (LPI), Zachary Williams (NCSU), Jaime Cordova (UW-Madison), Emmanuel Marcq (LATMOS/IPSL/UVSQ), Shannon Curry (UC Berkeley), Darby Dyar (PSI), Jason Rabinovitch (JPL), Pat Beauchamp (JPL), Jeff Balcerski (NASA Glenn), Kandis-Lea Jessup (SWRI), Kerrin Hensley (BU), Amanda Brecht, Ryan McCabe (Hampton), Tibor Kremic (NASA GRC), Alexey Martynov (LAV), Anastasia Kosenkova (LAV), Scott Guzewich (NASA GSFC), Gabriella Gilli (IA), Shaosui Xu (UC Berkeley), Lynn Carter (UA), Maxence Lefevre (Oxford), Kevin McGouldrick (UC Boulder), Kandis-Lea Jessup (SWRI), Colin Wilson (SSI), Scott King (Virginia Tech), Nader Haghighipour, (IfA, Hawaii), Zach Adam (UA), Vladimir Airapetian (NASA GSFC and American University), David A. Williams (ASU), Mikhail Zolotov (ASU), Brian Drouin (JPL), Tim Lichtenberg (Oxford), Joshua Krissansen-Totton (UC Santa Cruz), Joshua Pepper (Lehigh), Nathan Mayne (Exeter), Thomas Fauchez (NASA GSFC), James B. Garvin (NASA GSFC), Joe Renaud (NASA GSFC), David R. Ciardi (NExScI-Caltech/IPAC), Knicole Colón (NASA GSFC), Tyler D. Robinson (NAU), Sarah Rugheimer (Oxford), Edward Schwieterman (UC Riverside), Robin Wordsworth (Harvard), Aki Roberge (NASA GSFC), Dawn M. Gelino (NExScI-Caltech/IPAC), Timothy Lyons (UC Riverside), Wade Henning (University of Maryland, NASA GSFC), Evan L. Sneed (Penn State), Ravi Kopparapu (NASA GSFC), Jacob Lustig-Yaeger (UW), Devanshu Jha (MVJCE), Kathleen McIntyre (UCF), Jennifer G. Blank (NASA ARC), Jessica Noviello (ASU), Sukrit Ranjan (MIT), Gwen Hanley (Berkeley), Timothy Holt (USQ), Laura Schaefer (Stanford), Karalee K. Brugman (ASU), Matthew Tiscareno (SETI), Erika Kohler (NASA GSFC), Marshall J. Styczinski (UW), Nicolas Iro (Vienna), Indhu Varatharajan (DLR), Emily Martin (UC Santa Cruz), Joseph O'Rourke (ASU), Lynnae C. Quick (GSFC), Aditya Chopra (Groningen & ANU), Padma A. Yanamandra-Fisher (SSI), Jani Radebaugh (Brigham Young)

## Abstract

The key goals of the astrobiology community are to identify environments beyond Earth that may be habitable, and to search for signs of life in those environments. A fundamental aspect of understanding the limits of habitable environments and detectable signatures is the study of where such environments can occur. Thus, the need to study the creation, evolution, and frequency of environments hostile to habitability is an integral part of the astrobiology story. The study of these environments provides the opportunity to understand the bifurcation between habitable and uninhabitable conditions on planetary bodies. The archetype of such a planet is Earth's sibling planet, Venus, which provides a unique opportunity to explore the processes that created a completely uninhabitable environment and thus define the conditions that rule out bio-related signatures. We advocate a continued comprehensive study of our neighboring planet, to include models of early atmospheres, compositional abundances, and Venus-analog frequency analysis from current and future exoplanet data. Critically, new missions to Venus that provide in-situ data are necessary to address the major gaps in our current understanding, and to enable us to take the next steps in characterizing planetary habitability.

## 1. The Importance of Studying Venus to Exoplanetary Science

The prime focus of astrobiology research is the search for life elsewhere in the universe, and this search proceeds with the pragmatic methodology of looking for water and Earth-like conditions. In our solar system, Venus is the most Earth-like planet in terms of mass, size, and likely bulk composition, yet at some point in planetary history there was a bifurcation between the two: Earth has been continually habitable since the end-Hadean, whereas Venus became uninhabitable, likely due to ocean loss. Indeed, under this bifurcation scenario, Venus is the archetype for a world that transitioned from Earth-like conditions, through the inner edge of the Habitable Zone (HZ); thus it provides a natural laboratory to study the evolution of habitability for an Earth-sized planet. If we seek to understand habitability, proper understanding of the boundaries of the HZ are necessary: further study and development of our understanding of the evolution of Venus' environment are imperative. Furthermore, current and near-future exoplanet detection missions are biased towards close-in planets (see Section 4), so the most suitable targets for the *James Webb Space Telescope (JWST)* are more likely to be Venus-like planets than Earth-like planets. Incomplete understanding of the evolution of Venus' atmosphere and its present state will hinder the interpretation of these observations, motivating urgent further study.

## 2. The Current Venus Environment

Venus could be considered an "Earth-like" planet, because it has a similar size and bulk composition. However, it has a 92-bar atmosphere comprising 96.5% $CO_2$ and 3.5% $N_2$, and a surface temperature of 735 K. Venus' atmosphere is explained by a past runaway greenhouse (Walker 1975), which occurred when insolation exceeded the limit on outgoing thermal radiation from a moist atmosphere (Komabayashi 1967; Ingersoll 1969; Nakajima et al. 1992; Goldblatt & Watson 2012; Goldblatt et al. 2013), evaporating any oceans present. It is unclear whether such oceans condensed, then later evaporated (Way & Del Genio 2020), or never condensed after accretion (Hamano et al. 2013). In either case, water loss by hydrogen escape followed, evident in high D/H relative to Earth (Donahue 1982). Complete water loss would take a few hundred million years (Watson et al. 1981), but may have been throttled by oxygen accumulation (Wordsworth & Pierrehumbert 2013). Moreover, the Venusian nitrogen inventory is poorly constrained and may hold important clues to the atmospheric and mantle redox evolution (Wordsworth 2016). Notably, massive water loss during a runaway greenhouse has been suggested as producing substantial $O_2$ in exoplanet atmospheres (Wordsworth & Pierrehumbert 2014; Luger & Barnes 2015), but Venus serves as a counter-example to this. Hydration and oxidation of surface rocks (e.g., Matsui & Abe 1986) or top-of-atmosphere loss processes (Chassefière 1997; Collinson et al. 2016) are potential sinks for water. **Thus, Venus is an ideal laboratory to test hypotheses for abiotic oxygen production and loss processes.**

Cloud-top variations of $SO_2$ have been documented across several decades from *Pioneer Venus* to *Venus Express* observations (Marcq et al. 2012), implying a long-term atmospheric cycling mechanism, or possibly injections via volcanism. Recently, nine emissivity anomalies from compositional differences were identified by the Visible and Infrared Thermal Imaging Spectrometer (VIRTIS) aboard *Venux Express* as sites of potentially recent volcanism (Smrekar et al. 2010). There are purported lava flows associated with these anomalies estimated to be 2.5 million years old at most, and likely to be as young as 250,000 years old or even younger (Smrekar et al. 2010) based on expected weathering rates of freshly emplaced basalts. The emissivity anomalies sit atop regions of thin, elastic lithosphere according to *Magellan* gravity data, strengthening the volcanism interpretation. In 2015, additional evidence for active volcanism on Venus was uncovered with a new analysis of *Venus Express*' Venus Monitoring Camera (VMC) data. Four temporally variable surface hotspots were discovered at the Ganiki Chasma rift zone near volcanoes Ozza and Maat Montes (Shalygin et al., 2015), suggestive of present volcanic activity. However, interpreting these types of observations from above the cloud layer correctly is a challenge. The scattering footprint of radiation from the Venus surface escaping through the cloud deck is about 100 km$^2$, so smaller areas of increased thermal emission are smeared out. **Venus clearly has a complex interaction between the interior, surface, and atmosphere that will play a key role in the diagnosis of exoplanetary atmospheres.**

## 3. Critical questions: The Need to Understand Earth's Twin
Many significant questions remain regarding the current state of Venus, leaving major gaps in our understanding of the evolution of silicate planets, including the future evolution of Earth (Kane et al. 2019), as well as our understanding of terrestrial exoplanets. Some of those major questions include:

- Did Venus have a habitable period (e.g. Way et al. 2016)? That is, did Venus ever cool from a syn-accretionary runaway greenhouse?
- Where did the water go? Was hydrogen loss and abiotic oxygen production rampant, or did surface hydration dominate? Were crustal rocks oxidized by water to form ferric oxides and H escaped?
- What has the history of tectonics, volatile cycling, and volcanic resurfacing been (Ivanov & Head 2011)? When did Venus enter its present stagnant-lid regime? How much, if any, volatile recycling occurs today?
- What is the precise structure of Venus' interior, why does Venus have a negligible magnetic field, and how has the lack of a strong magnetosphere affected the erosion of the Venusian atmosphere (Nimmo 2002; Jacobson et al. 2017).
- What is the composition of and atmospheric chemistry within the Venusian middle and deep atmosphere and how does the atmosphere interact with the surface and influence its mineralogy (Zolotov 2018)? Are there regions in the present atmosphere that could support life (Dartnell et al. 2015; Limaye et al. 2018)?

A key focus of current and future exoplanet science is the measurement and modeling of terrestrial atmospheres (Turbet et al. 2016) and their escape (Dong et al. 2020). The method of transmission spectroscopy during planetary transit is the primary means through which such measurements can be obtained (Seager & Sasselov 2000), and will be used to investigate upper atmospheric compositions from *JWST* observations (Lincowski et al. 2018; Lustig-Yaeger et al. 2019). Interpretation of these data are sensitive to the composition, chemistry, and dynamics of the deeper atmosphere which is largely opaque at most wavelengths. Measurements of isotopic ratios in the Venusian atmosphere provide critical information regarding the history of atmospheric loss (Gillmann et al. 2009) and enhanced D/H ratios may even be accessible in terrestrial exoplanet atmospheres with *JWST* (Lincowski et al. 2019). **It is therefore imperative to obtain additional in situ data for terrestrial atmospheres within our solar system, particularly for a diverse range of atmospheric chemistries.**

Venus accounts for 40% of the mass of terrestrial planets in our solar system, yet even fundamental parameters, such as the relative size of its core, are largely unknown. As we expand the scope of planetary science to include those planets around other stars, the lack of measurements for basic planetary properties of Venus such as its moment of inertia, core size and state, seismic velocity and density variations with depth, and thermal profile hinders our ability to compare the potential uniqueness of the Earth and our solar system to other planetary systems.

Furthermore, determining the relative abundances of Venus' refractory elements can greatly inform us of the degree of mixing of planetesimals within the critical zone in the disk: where terrestrial planets form. If Venus' relative refractory ratios are reflected in the size of its core, we gain a key benchmark in future studies of how our solar system formed by constraining even this simple parameter for the second planet. In turn, we will be greatly aided in our studies of exoplanets, where stellar composition may set the initial compositional gradient of planetesimals within the disk but where the degree of mixing remains an elusive, unconstrained parameter.

## 4. A Plethora of Venus Analogs

The inner and outer boundaries of the HZ for various main sequence stars have been estimated with climate models, such as those by Kasting et al. (1993), and more recently by Kopparapu et al. (2013, 2014). An important aspect of these HZ calculations is that they provide a means to estimate the fraction of stars with Earth-size planets in the HZ, what we term "eta-Earth". Much of the recent calculations for eta-Earth utilize *Kepler* results since these provide a large sample of terrestrial size objects from which to perform meaningful statistical analyses (Dressing & Charbonneau 2013, 2015; Kopparapu 2013; Petigura et al. 2013).

The transit method has a dramatic bias towards the detection of planets that are closer to their host star (Kane & von Braun 2008). Additionally, a shorter orbital period results in an increased signal-to-noise ratio of the transit signature because of the increased number of transits observed. Consequently, *Kepler* has preferentially detected planets interior to the HZ that are more likely to be potential Venus analogs than Earth analogs (Kane et al. 2018). Since the prospect and timing of the divergent evolutions of the Earth and Venus atmospheres is a critical component for understanding Earth's habitability, the frequency of Venus analogs ("eta-Venus") is also important to quantify.

Kane et al. (2014) defined the "Venus Zone" (VZ) as a target selection tool to identify terrestrial planets where the atmosphere could potentially be pushed into a runaway greenhouse to yield surface conditions similar to those of Venus. The below figure shows the VZ (red) and HZ (blue) for stars of different temperatures. The outer boundary of the VZ is the "runaway greenhouse" line which is calculated using climate models of Earth's atmosphere. The inner boundary (red dashed line) is estimated based on where the stellar radiation from the star would cause complete atmospheric erosion. The pictures of Venus shown in this region represent planet candidates detected by *Kepler*. Kane et al. (2014) calculated an occurrence rate for VZ terrestrial planets of 32% for low-mass stars and 45% for Sun-like stars. Note however that, as for the HZ, the boundaries of the VZ should be considered a testable hypothesis since a runaway greenhouse could occur beyond the calculated boundary (Hamano et al. 2013; Foley 2015).

The prevalence of Venus analogs will continue to be relevant in the era of the *Transiting Exoplanet Survey Satellite (TESS)* mission, as hundreds of terrestrial planets orbiting bright host stars are expected to be detected (Sullivan et al. 2015; Barclay et al. 2018). These detections will provide key opportunities for transmission spectroscopy follow-up observations with *JWST* (Kempton et al. 2018), amongst other facilities and ground-based telescopes. Such efforts to identify key atmospheric abundances for terrestrial planets will face the challenge of distinguishing between possible Venus and Earth-like surface conditions (Lustig-Yaeger et al. 2019; Ostberg & Kane 2019). **Discerning the actual occurrence of Venus analogs will help us to decode why the atmosphere of Venus is so radically different to its sibling planet, Earth.**

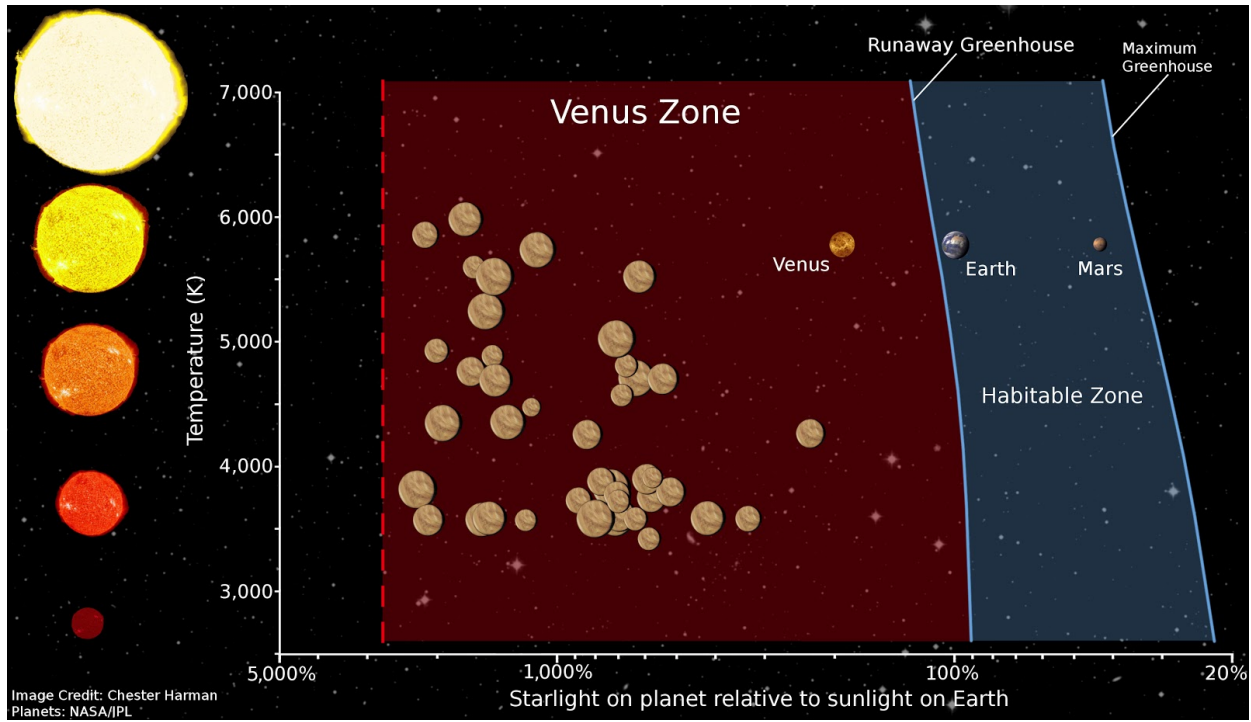

*The extent of the Venus Zone as a function of host star temperature and incident flux, where solar system planets and terrestrial Kepler candidates are shown. Credit: Chester Harman.*

## 5. The Path Forward

The only in-situ terrestrial planet data available to us are here in our solar system. Thus, it is imperative that we gather improved information on Venus to aid in modeling planetary atmospheres, surfaces, and interiors. The greatest advances in studies of Venus will come from a better understanding of the top-level questions described in Section 3 for which a series of missions - at multiple cost scales - could address parts thereof. Atmospheric modeling of exoplanets is of critical importance and an improved sampling of pressure, temperature, composition, and dynamics of the Venusian atmosphere as a function of latitude and altitude would aid enormously in our ability to study exoplanetary atmospheres. In particular, new direct measurements of D/H within and below the clouds are needed to better constrain the volume of water throughout Venus' history. Combined with improved bounds on the D/H ratio, isotopic measurements in the atmosphere would yield insights into the origins and fate of the Venusian atmosphere. A descent probe or lander to the surface (as a Discovery, New Frontiers, or part of a larger flagship mission) could make vital new measurements of atmospheric structure and D/H ratios, as well as noble gas abundances and isotopic ratios. Such a mission could also provide first-ever measurements of the deepest Venus atmosphere. Aerial platforms, such as balloons complement the vertical probe profiles by providing 2-D coverage of the cloud region. Orbiting radar missions to determine Venus' detailed geological evolution, current level of activity, and indications of key geodynamic changes with time, are also essential. These fundamental measurements would stimulate progress on multiple fronts, and vastly improve our understanding of both modern Venus and the history of the planet, including the path it has taken to its modern state, which will help constrain the fundamental ocean and atmospheric loss processes that brought it to its modern state, and that likely also affect terrestrial exoplanets. The inclusion of a seismometer on future landers or aerial platforms to measure moment of inertia, will provide new critical but necessary information about the Venusian interior, with which we will be able to expand our inferred knowledge of *any* exoplanet system. Some of these science goals would be accomplished by recently selected missions for further design studies, including a Venus Flagship Mission (VFM) concept study (PI: Gilmore), and VERITAS (PI: Smrekar) and DAVINCI+ (PI: Garvin), both shortlisted for NASA's Discovery Program. Given the trajectory of exoplanetary science towards exoplanet atmospheric characterization and the expected prevalence of Venus analog environments, the study of Venus represents the highest priority synergistic target between the planetary science and exoplanet communities. **Therefore it is a matter of high priority that crucial pressure, temperature, and compositional atmospheric measurements of the nearest Earth-size planet are obtained.**